\documentclass[12pt]{report}
\usepackage{graphicx}

\begin{document}

\title{Notes on the data analysis in high-energy astrophysics}

\author{Luigi Foschini\\
\emph{INAF - Osservatorio Astronomico di Brera}\\
\emph{Via E. Bianchi, 46 - 23807 - MERATE (LC), Italy}\\
email: \texttt{luigi.foschini@brera.inaf.it}\\
Home page: \texttt{http://www.brera.inaf.it/utenti/foschini/}
}

\maketitle

\newpage

\begin{center}
\textbf{Foreword}
\end{center}

These notes were originally prepared as additional material for the lessons I have given at the summer school 
\emph{``Gamma-ray Astrophysics and Multifrequency: Data analysis and astroparticle problems''}(\footnote{\texttt{http://glastweb.pg.infn.it/school2006/}.}), organized by the Department of Physics of the University of Perugia (Italy) on July 3-7, 2006. The necessarily limited time of the lessons forced to a drastic selection of the topics and, therefore, I have thought it was useful to complete the slides of the presentation with the notes you find in these few pages. 

These notes are a kind of ``link'' between the theoretical approach of the University lessons and the practice of the real use. During these years, I have seen that this material is useful for students that are beginning their thesis on research topics based on the use and the interpretation of data from X- and $\gamma-$ray satellites.  Therefore, I have thought to update the content of these notes, to translate them into English and to post on \texttt{arXiv}, in order to make them available to a larger public, hoping to give a useful help to other beginners. 

Since the public is basically composed of graduating students, I have assumed a well-grounded knowledge of basic astrophysics, the principles of instrumentation for physical sciences and the statistical analysis. Therefore, I have focused my notes on the practical issues and given some bibliographic references, just some advice, just to draw the attention on some points that are too much detailed for a university lesson. Nevertheless, the bibliography should be also a good starting point for those who want to further explore these topics.

\vskip 12pt
Merate (LC), October 12, 2009

\vskip 12pt
\begin{center}
\emph{Luigi Foschini}
\end{center}

\newpage

\section{Basic Concepts}

\subsection{Definitions}
It is possible to divide the analysis of astrophysical data into two broad parts: the first one, based on the analysis of the images, has the aim of measuring
the coordinates of the sources on the sky, their spatial structure (if extended or point-like), and the photometry of the sources. The second part, which is
based on the study of the emitted electromagnetic spectra, is dedicated to the study of the flux and the energy of the emission or absorption lines and the 
characteristics of the continuum. 

The measurement process introduces several effects that contribute to change, in a negative way, the original spatial structure and spectrum of the source. Moreover, there are external disturbances (e.g. cosmic and instrumental background) that worsen the instruments performance. Therefore, it is necessary to make corrections to the data in order to obtain an image or a spectrum that is as much as possible in agreement with the original one. Having a reliable information from the observations is the basis for the interpretation and development of theories.

Although in astrophysics there are different types of instruments, each of them with its performance and problems, from a conceptual point of view the data analysis is nothing else than solving an \emph{inverse problem with distributions} (or generalized functions). For example, let us to consider the simplest case, with one dimension. It is necessary to solve the following Fredholm equation of the first kind:

\begin{equation}
\phi(x)=\int \psi(y)K(x,y)dz
\label{fred1}
\end{equation}

\noindent where $\phi$ is the observed density of probability and $\psi$ is the corresponding original density of probability that we want to restore. $K$ is the kernel of the integral equation and represents the measurement process. This includes both the effects of the measurement errors and the external effects altering the observed distribution. The kernel is named \emph{point--spread function} (PSF) for the images and \emph{line--spread function} (LSF) for the spectra. $\phi$, $\psi$, and $K$ are all non negative functions, because are derived from the density of incident photons.

Therefore, the fundamental problem of the data analysis is to clean the PSF or the LSF from all the measurement errors (including those of calibration) before inverting
the integral equation. After this cleaning process, the kernel \emph{should} contain only the statistical errors intrinsic to the measurement. For example, in the case of
a gaussian distribution with variance $\sigma^2$, the final kernel is:

\begin{equation}
K(x,y)=\frac{1}{\sigma \sqrt{2\pi}}\exp{\frac{-(x-y)^2}{2\sigma^2}}
\label{fred2}
\end{equation}

\subsection{Data processing}
A system for the processing of astronomical data should include at least three levels, which in turn can be divided into sublevels, depending on the instrumentation
used in the observations.

The first level (\textbf{Level 0}) is the \emph{preprocessing}: the telemetry coming from the satellite is uncompressed and converted in the standard format FITS 
(\emph{Flexible Image Transportation System}). No other operation is performed on the data, which are therefore named ``raw''. The basic data structure adopted for
the instruments onboard high-energy astrophysics satellites is a list of events, where each row contains at least the onboard time, the detector position ($y,z$) where the
interaction occurred and the channel of energy. Sometimes, depending on the need, it is possible to perform an early processing onboard the satellite: for example,
the data of PICsIT, the high-energy detector of the IBIS telescope onboard the \emph{INTEGRAL} satellite, are equalized and integrated onboard, in order to save
telemetry. On the other hand, this has the drawback of reducing the possibility to correct -- with some dedicated specific software -- the data downloaded to ground.
When possible, it is always preferable to download the data without performing any onboard processing, so to have always the possibility to re-process the raw data,
by applying new corrections, calibrations and improvements that can be developed later. 

Once the basic data structure has been created and the raw data are available, it is possible to begin with the next stage of the processing, which is the correction
of the data (\textbf{Level 1}). At this stage, all those instrument-dependent corrections necessary to clean the kernel of the Eq.~(\ref{fred1}) are applied. The most
important are:

\begin{enumerate}
\item the intrinsic detector \emph{deadtime} that is the time during which it is not possible to record any event, since the detector is already busy with another event;
therefore, the number of recorded counts has to be increased to take into account these periods where, even in presence of a photon flux, the detector cannot collect them;

\item each photon with energy $E$ is recorded in a certain energy channel; the conversion from the channel to energy (\emph{gain}) generally depends on the
temperature and therefore it is necessary to compensate any gain changes with respect to the nominal value;

\item in the CCD detectors, it is necessary to take into account the (\emph{rise time}) and charge losses;

\item filtering of the events due to the Good Time Intervals (GTI); these are time periods where the instrument is nominally working and the satellite is correctly
pointing the planned source; conversely, the bad time intervals occur when the instrument has some parameter (\emph{housekeeping}) out of its nominal range (e.g. an anomalously high temperature) or the satellite is passing in a spatial region with high background (e.g. the South Atlantic Anomaly);

\item correction of non uniformities of the detector or the deformation of the PSF in focussing telescopes (\emph{vignetting}), subtraction of the cosmic and instrumental background.
\end{enumerate}

Now, we are ready to pass to the third level (\textbf{Level 2}), where images, spectra and light curves are generated. Images should be supplied with 
celestial coordinates, starting from the reference values given by the \emph{star tracker}, a small optical telescope used to measure the reference coordinates of the main satellite axis. Since the instruments are never aligned with the star tracker, it is necessary to perform a rototraslation of the coordinates array.

To project the image of the detector on the sky, there are several types of projections that can be used depending on the cases. The most diffuse projection is the
gnomonic or tangential one (TAN), but it is useful for small fields of view distant from the poles. Other projections (ARC, AIT, STG and so on...) are used to by-pass these limits. Aitoff (AIT) projection is generally used for all-sky maps, while the stereographic (STG) is adopted for large fields of view. The standard products of the image analysis can be lists of sources (source detection), with coordinates, flux (in counts per second or physical units) and the significance or error of measurement.

The sources count spectra (in units of counts per second per energy bin) have to be convolved with the response of the instrument. Two files are important for this purpose:
the Auxiliary (or Ancillary) Response File (ARF) and the Redistribution Matrix File (RMF). The ARF contains the information of the effective area of the instrument as a function of the photon energy, that is the geometric area of the detector convolved with the response of the optics, the possible effects of the degradation of the PSF, the quantum efficiency, and so on... If you convolve the count spectrum with the only ARF, then you have the response of a detector with infinite energy resolution. 

To stop dreaming and return to ground, it is necessary to use the RMF, which contains the necessary information to know how photons are scattered in the energy channels, 
in order to convert the detector counts into photons and then into physical units.

Sometimes, the information contained in the ARF and RMF files are merged into a single file (e.g. IRF, Instrument Response Function for the LAT detector onboard \emph{Fermi}). 

\vskip 12pt
\textbf{To know more:}

\begin{itemize}
\item Definition of the FITS format: 
\subitem Wells D.C., Greisen E.W., Harten R.H.: ``FITS - a Flexible Image Transport System'', \emph{Astronomy and Astrophysics Supplement Series},  \textbf{44}, (1981), 363-370.

\item Web site: \texttt{http://heasarc.gsfc.nasa.gov/docs/heasarc/fits.html}.

\item The FITS format is approved by the \emph{International Astronomical Union} (IAU), which has set up a dedicated working group: \\
\texttt{http://fits.gsfc.nasa.gov/iaufwg/iaufwg.html}.

\item World Coordinate System (WCS), where are described the different types of sky projections in astronomy: 
\subitem Greisen E.W. \& Calabretta M.R.: ``Representations of world coordinates in FITS'', \emph{Astronomy and Astrophysics}, \textbf{395}, (2002), 1061-1075; 
\subitem Calabretta M.R. \& Greisen E.W.:  ``Representations of celestial coordinates in FITS'', \emph{Astronomy and Astrophysics}, \textbf{395}, (2002) 1077-1122.

\item WCS web site: \texttt{http://www.atnf.csiro.au/people/mcalabre/WCS/index.html}.

\item Web site with libraries and tools for WCS:\\ 
\texttt{http://tdc-www.harvard.edu/software/wcstools/index.html}.

\end{itemize}

\newpage

\section{Instruments}
\begin{figure}
\centering
\includegraphics[scale=0.65]{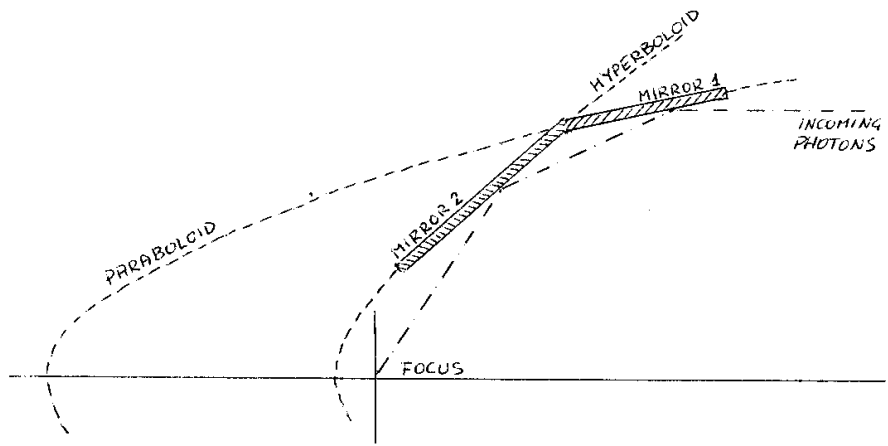}
\caption{Giacconi's optics for X-ray telescopes.}
\label{fig:giacconi}
\end{figure}

The best way to build astronomical instruments is to \textbf{focus} the photons from cosmic sources on a restricted area of a position sensitive detector. However, as
the energy of the incoming photon increases, the refraction index approaches 1 and therefore it is more and more difficult to deviate it according to the
requirements of the focalization. It is worth noting that even though the index approaches 1, \emph{it is not 1}, and it is possible to build optics for X-ray 
astronomy by using the grazing incidence techniques developed and applied to astronomy by R. Giacconi (Nobel Prize for Physics in 2002). A paraboloidal mirror with a coaxial and confocal hyperboloid one allows to efficiently focus grazing X-ray photons (Fig.~\ref{fig:giacconi}). The present day technologies allow us to focus photons with energies up to $\approx 10$~keV (e.g., \emph{Chandra}, \emph{XMM-Newton/EPIC}, \emph{Swift/XRT}). There are active projects aimed to focus photons up to $\approx 100$~keV. However, for the moment, the only way to build instruments operating at energies $>10$~keV are collimators and coded masks. 

The \textbf{collimation} is the simplest type of optics: it is basically a shield of absorbing material surrounding a detector, except for a small zone, corresponding to
a sky region of some degree or arcminute size. Also the working principle is quite simple and easy: photons are collected from a region of the sky without sources (empty field), in order to measure the cosmic background, and are subtracted from those collected in a region where there is the observed source plus obviously the background. This method is named ``on/off''. Its main advantage is to have a sensitivity greater than imaging instruments, but the weakness is the possibility of contamination by nearby sources, which increases as the field-of-view (FOV) widens. Instruments of this type, because of their simplicity, have been used since the dawn of the high-energy astrophysics: \emph{Uhuru}, which can be defined the first true X-ray astrophysics satellite, had two collimated proportional counters with FOVs of $0^{\circ}.52\times 0^{\circ}.52$ and $5^{\circ}.2\times 5^{\circ}.2$.
Presently, two satellites are in orbit working in the hard X-ray energy band with collimation instruments:  \emph{Rossi X-ray Timing Explorer} (launched in $1995$, energy band $2-250$~keV) and \emph{Suzaku} (launched in $2005$, energy band $10-600$~keV).

\begin{figure}
\centering
\includegraphics[scale=0.3]{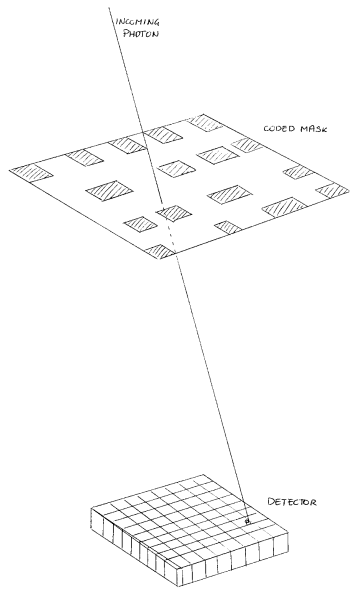}
\caption{Concept of coded masks.}
\label{fig:cmask}
\end{figure}

The \textbf{coded masks} are optics for hard X-ray that allow imaging. The flux of incoming photons is spatially modulated according to a mathematical sequence, selected
on the basis of the type of structures that are to be emphasized. The basic principle is to shield part of the incoming photons and to allow another part to reach the detector (Fig.~\ref{fig:cmask}). Therefore, a shadowgram is created on the detector, generated by the convolution of the photon flux with the mask and from which it is possible to restore the original image of the sky by using the deconvolution operation. This technique is analyzed with some detail in Section 4. 

Although, coded mask instrument were used almost in the past, today they are living a second youth thanks to the satellites \emph{INTEGRAL} (launched in $2002$, with 3 coded mask instruments covering the $0.003-10$~MeV energy band) and \emph{Swift} (launched in $2004$, with 1 coded mask instrument in the $15-200$~keV energy band). These instruments are less sensitive than those with collimators, but have the advantage of generating images, thus reducing the problem of \emph{source confusion}, particularly important in Galactic astrophysics. Moreover, this type of instruments guarantees an almost uniform response over a wide FOV, which is essential to monitor very large regions of the sky searching for GRB or outbursts from X-ray binaries or blazars. 

\begin{figure}
\centering
\includegraphics[angle=270,scale=0.65]{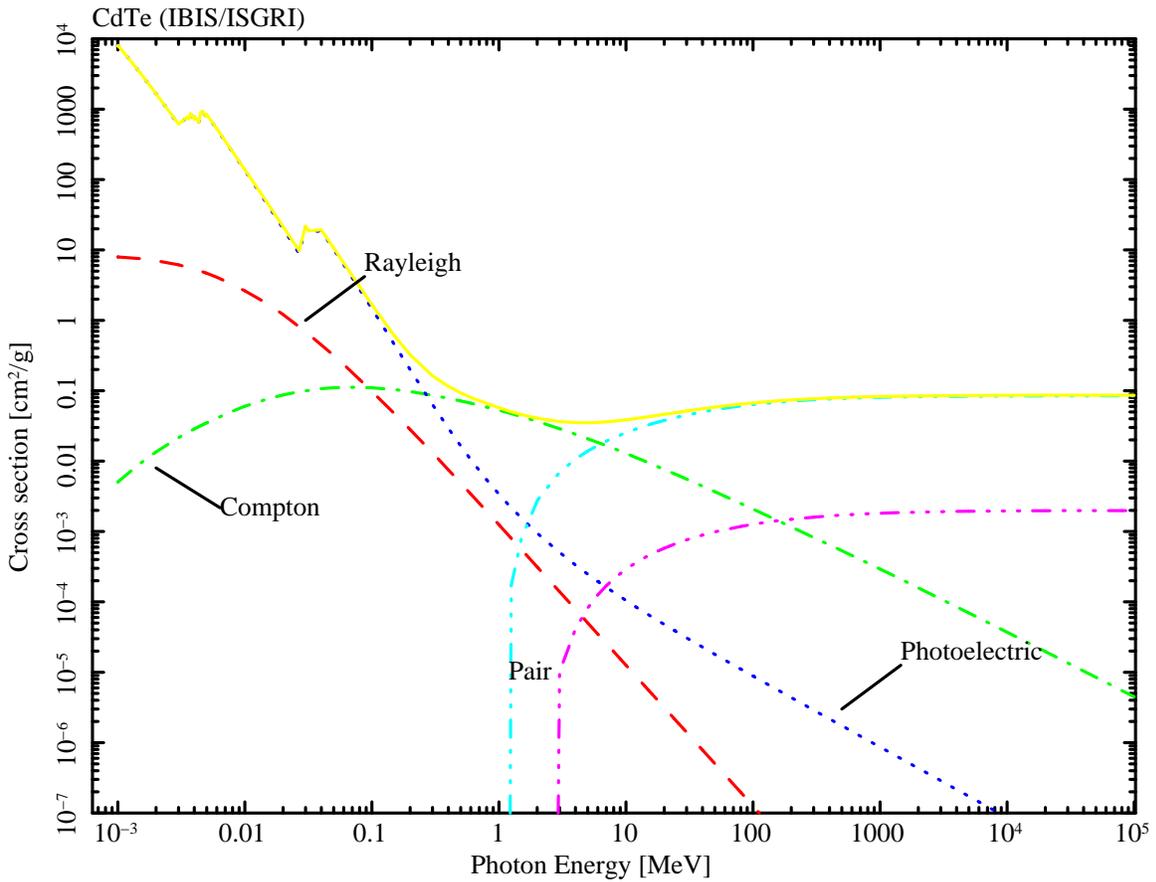}
\caption{Photon absorption cross section for CdTe (used for the IBIS/ISGRI detector onboard \emph{INTEGRAL}). The dashed lines represents the coherent scattering (Rayleigh), while the dot-dashed line is for the Compton scattering. The dotted line is the photoelectric absorption and the two lines triple dot-dashed indicate the pair production by electron or nuclear field. The continuum line is the sum of all the processes.}
\label{fig:cdte}
\end{figure}

These optics have to work together with proper detectors. X-ray photons are better absorbed by high-Z materials, as shown in Fig.~\ref{fig:cdte}, where it is displayed the photon absorption cross sections for different processes in the CdTe. This material has been used to build the IBIS/ISGRI detector onboard \emph{INTEGRAL} and a slightly different version - CdZnTe - has been used for BAT onboard \emph{Swift}. These detectors are the high-energy version of the CCD. At greater energies, hundreds of keV to a few MeV, the detectors are built with CsI or NaI (e.g. IBIS/PICsIT onboard \emph{INTEGRAL} is made of CsI), but the germanium is adopted if it is necessary a high energy resolution (e.g. SPI onboard {INTEGRAL}).

\begin{figure}
\centering
\includegraphics[scale=0.3,clip,trim = 0 0 0 40]{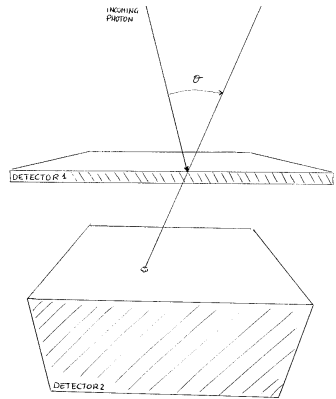}
\caption{Concept of Compton telescope. In the case of IBIS, ISGRI is the Detector 1, while PICsIT is the Detector 2.}
\label{fig:compton}
\end{figure}

As the energy of the photon increases, in the $\gamma-$ray energy band, when other processes become dominant (Compton effect and pair production), it is no more possible to separate the optics from the detector. In the MeV energy range, the \textbf{Compton effect} hampers any tentative to shield efficiently the detector, making thus quite difficult to reach high sensitivities. Nevertheless, some instruments have been developed to work using the Compton effect as operating principle. These instruments are made of two detectors: the incoming photon is scattered by the first detector and then absorbed by the second one (Fig.~\ref{fig:compton}). Knowing the energy deposits in the two detectors, it is possible to calculate a range of directions of the incoming photons. Although the sensitivity remains low with respect to instruments in other energy ranges, the selection of photons interacting with both detectors allows a drastic reduction of the background and thus an improvement of the performance. The poor angular resolution (a few degrees) can be improved by using coded masks (up to tens of arcminutes). Obviously, this solution is applicable at energies below $\approx 1$~MeV, where the coded masks are still effective (at greater energies also these optics become mostly transparent, thus loosing their ability to code the incoming flux). Examples of Compton telescopes are COMPTEL ($1-30$~MeV) onboard the \emph{Compton Gamma-Ray Observatory} (CGRO, launched in $1991$ and fell into the ocean in $2000$) and IBIS onboard \emph{INTEGRAL}, when its two detectors ISGRI and PICsIT work together ($\approx 0.2-1$~MeV). 

\begin{figure}
\centering
\includegraphics[scale=0.7,clip,trim = 0 50 20 20]{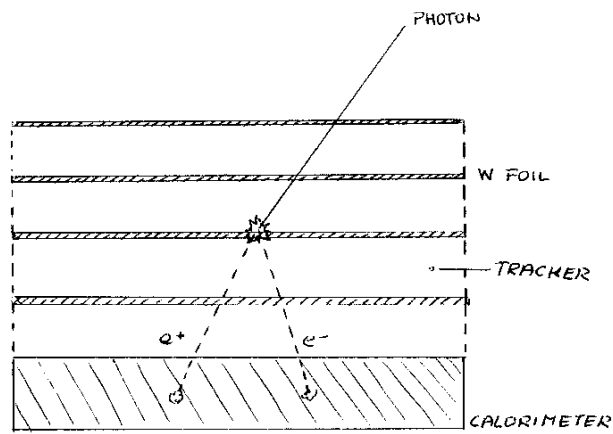}
\caption{Concept of pair conversion telescope.}
\label{fig:pair}
\end{figure}

In the GeV energy range, another process starts dominating, the \textbf{pair production}, thus making necessary to change again the concept of the telescope (Fig.~\ref{fig:pair}). In this case, the instrument is built with foils made of high-Z elements (e.g. W), having a high cross section for photon pair production, with a position sensitive detector in the middle (spark chambers in the past; silicon detectors today). After the creation of a pair in a W foil, the generated $e^+e^-$ pass through the silicon detector, which allows to track the path of the particles. A calorimeter on the bottom of the detector, allows to measure the energy with high precision. With the information on the trajectories of the particles and their deposited energy, it is possible to restore the direction and energy of the incoming photon. The first successful experiment of this type was EGRET onboard \emph{CGRO}, which worked in the $0.1-30$~GeV energy range. Today, two of such instruments are active: LAT onboard \emph{Fermi} (launched in $2008$, energy range $0.1-300$~GeV) and GRID onboard \emph{AGILE} (launched in $2007$, energy range $0.1-30$~GeV).

\vskip 12pt
\textbf{To know more:}

\begin{itemize}
\item Basic information and description of all the high-energy astrophysics missions (from the beginning to the forthcoming projects) can be found at: \\
\texttt{http://heasarc.gsfc.nasa.gov/docs/heasarc/missions/alphabet.html}.

\item The principles of the interaction of radiation with the matter, detectors, and materials can be found in the very good text book: 
\subitem W.R. Leo: ``Techniques for Nuclear and Particle Physics Experiments'' (Springer-Verlag, 1994).

\item It is not possible to think about focusing techniques for X-ray astronomy without citing the work by:
\subitem R. Giacconi, W.P. Reidy, G.S. Vaiana, L.P. Van Speybroeck, T.F. Zehnpfenning: ``Grazing-incidence telescopes for X-ray astronomy'', \emph{Space Science Reviews}, \textbf{9}, (1969), 3-57.

\item IBIS used as Compton telescope is described in:
\subitem M. Forot, P. Laurent, F. Lebrun, O. Limousin: ``Compton telescope with a coded aperture mask: imaging with the \emph{INTEGRAL}/IBIS Compton mode'', \emph{The Astrophysical Journal}, \textbf{668}, (2007), 1259-1265.

\item A description of the LAT instrument onboard \emph{Fermi} can be read in: 
\subitem W.B. Atwood et al.: ``The Large Area Telescope on the \emph{Fermi Gamma-ray Space Telescope} mission'', \emph{The Astrophysical Journal}, \textbf{697}, (2009), 1071-1102.

\end{itemize}

\newpage

\section{Image processing (with specific reference to coded mask instruments)}
From a physical point of view, the cosmic radiation interacts with the optics of the telescopes and deposits its energy in the detector. From a mathematical point of view, the sky map is convolved with the instrument response (PSF), as already outlined in Section 1 (we are now passing from hardware to software). 

The convolution of two real functions $f$ and $g$ is defined as:

\begin{equation}
(f*g)(x)=\int g(x-y)f(y)dy
\label{convol1}
\end{equation}

that is just another version of Eq.~(\ref{fred1}), where $\phi(x)$ is the convolution $(f*g)(x)$, $\psi(y)$ is $f(y)$ and the kernel $K(x,y)$ is $g(x-y)$, i.e. the PSF. 
The convolution operation can be thought as a ``moving average'' of $f$ (sky map) weighted by $g$ (PSF). 

However, in the physical reality, we have to deal with finite distributions (the counts in the pixels of a detector with finite dimensions) and, therefore, we have to substitute integrals with summations and distributions with matrices. Let us write with $F$ and $G$ two matrices with dimensions $M\times N$. Then, the convolution is now:

\begin{equation}
(F*G)_{mn}=\sum_{k=0}^{M-1}\sum_{l=0}^{N-1} G_{kl}F_{m-k,n-l}
\label{convol2}
\end{equation}

One interesting property of the convolution operator is that is transformed into a simple multiplication if we convert Eq.~(\ref{convol2}) to the frequency space, that is we perform a Fourier transform. Indicating with $\tilde{F}$ and $\tilde{G}$ the Fourier transform of the matrices $F$ and $G$, respectively, then the Eq.~(\ref{convol2}) changes to:

\begin{equation}
(F*G)_{mn}=MN\tilde{F}_{uv}\tilde{G}_{uv}
\label{convol3}
\end{equation}

Now, to restore the original sky map it is \emph{sufficient} to invert the Eq.~(\ref{convol3}) and then return to the space domain. The problem is that the matrix to be inverted can have no inverse. In other words, it is said that the problem is \emph{ill posed} and it is necessary to use regularization techniques. In the case of coded masks, it is possible to adopt a different way, the spatial correlation that is quite similar to the convolution, but it is defined as:

\begin{equation}
(F\otimes G)_{mn}=\sum_{k=0}^{M-1}\sum_{l=0}^{N-1} G_{kl}F_{m+k,n+l}
\label{correlaz}
\end{equation}

Comparing Eq.~(\ref{convol2}) with the Eq.~(\ref{correlaz}), it is possible to note that the convolution is a correlation plus a reflection. Therefore, once known this difference, there is no problem in using the former or the latter. Again, in this case, it is possible to perform a Fourier transform:

\begin{equation}
(F\otimes G)_{mn}=MN\tilde{F}_{uv}\tilde{G}_{uv}^{*}
\label{correlaz2}
\end{equation}

where $\tilde{G}^*$ is the complex conjugate matrix of $\tilde{G}$.

In practice, to deconvolve a sky region by using the correlation, it is necessary to build a decoding array. Let us to indicate with the symbol $D$ (data) the shadowgram projected by the coded mask onto the detector, $B$ the background, $M$ the array describing the mask (made with $1$ for open pixels and $0$ for closed pixels), and $S$ (sky) the observed region of the sky, then $D=S\otimes M+B$. In order to deconvolve with the correlation, it is necessary to search for the decoding array $G$, build so that $M\otimes G=\delta$, where $\delta$ is the array corresponding to the Dirac function. Now, the reconstructed sky $S'$ can be calculated with this simple operation:

\begin{equation}
S'= D\otimes G = S\otimes M \otimes G + B\otimes G = S\otimes \delta + B\otimes G = S + B\otimes G
\label{correlaz3}
\end{equation}

It is worth noting that the sky map $S'$ is different from the original sky map $S$ only for the term $B\otimes G$ (background), which in turn has to be subtracted with the proper techniques.

\begin{figure}
\centering
\includegraphics[scale=0.4,clip,trim = 0 0 0 0]{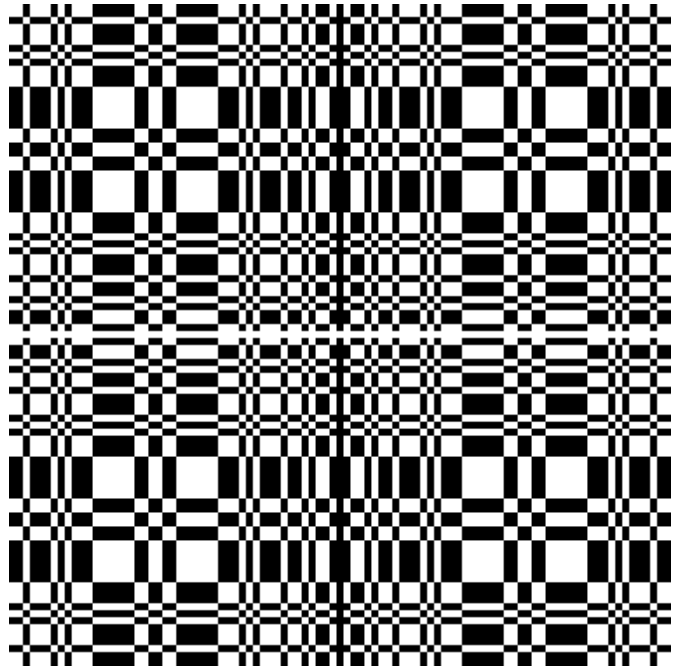}
\caption{MURA mask for the IBIS telescope.}
\label{fig:ibismask}
\end{figure}

In the case of the Modified Uniformly Redundant Array (MURA) mask (Fig.~\ref{fig:ibismask}), used by IBIS, the decoding array $G$ is $2M-1$, i.e. one element of $G$ is equal to $1$ when one element of $M$ is equal to $1$. If the latter is equal to $0$, then $G=-1$.

\vskip 12pt
\textbf{To know more:}

\begin{itemize}
\item A good and detailed book on the processing of images is
\subitem B. J\"ahne: ``Practical Handbook on Image Processing for Scientific Applications'' (CRC Press, 1997).

\item A detailed description of the data analysis of IBIS onboard \emph{INTEGRAL} can be read on the papers: 
\subitem A. Goldwurm et al., ``The INTEGRAL/IBIS Scientific Data Analysis'', \emph{Astronomy and Astrophysics}, \textbf{441}, (2003), L223-L229; 
\subitem A. Gros et al.: ``The INTEGRAL IBIS/ISGRI System Point Spread Function and Source Location Accuracy'', \emph{Astronomy and Astrophysics} 441, (2003), L179-L183. 

\item The MURA mask, adopted by IBIS, is described in the work by: 
\subitem S.R. Gottesman \& E.E. Fenimore, ``New family of binary arrays for coded aperture imaging'', \emph{Applied Optics}, \textbf{28}, (1989), 4344-4352.

\item In the case of BAT, onboard \emph{Swift}, see: 
\subitem S.D. Barthelmy et al., ``The Burst Alert Telescope (BAT) on the Swift Midex Mission'', \emph{Space Science Reviews}, \textbf{120}, (2005), 143-164.

\item The reviews by Gerry Skinner are very useful and witty. Just to cite a couple of them: 
\subitem ``Imaging with coded-aperture masks'', \emph{Nuclear Instruments and Methods in Physics Research}, \textbf{221}, (1984), 33-40; 
\subitem ``Coded mask imagers when to use them -- and when not'', \emph{New Astronomy Reviews}, \textbf{48}, (2004), 205-208.

\item Last, but not least, there is the web page on coded mask developed by Jean in't Zand: \\
\texttt{http://universe.gsfc.nasa.gov/cai/}.

\end{itemize}

\newpage

\section{Analysis of errors and other processes altering the measurement}
The statistical analysis of measurement errors is widely dealt in several treatises and textbooks (see the bibliography at the end of this section): therefore, I will not repeat here these concepts and I will draw your attention to some practical issues. 

The \emph{vexata quaestio} of the statistical analysis in astrophysics is the \textbf{significance}: when is it acceptable? Well, basically there is no clear-cut definition and it depends on several factors. Astrophysics is not laboratory physics, when you can control almost all the factors of the experiment and then $5\sigma$ ($99.99994$\% confidence level) is a ``must''. Specifically in space astrophysics, you have to face with the fact that your instrument is deployed into orbit and therefore you have no possibility to fully control or update it\footnote{Except for the \emph{Hubble Space Telescope}, which has a modular design, so that it is possible to upgrade the instruments in orbit. The spectacular \emph{Space Shuttle} missions aimed to this purpose, with astronauts working in orbit on \emph{HST}, are well known to the public.}. In addition, you have often to deal with a few photons, perhaps due to a low sensitivity of your instrument or a very short exposure or a high-background period during your observation and so on. Last, but not least, \emph{you cannot repeat the experiment!} Therefore, sometimes, it is necessary to relax the requirements on the significance to lower values, like $3\sigma$ ($99.73$\% c.l.) or even $2\sigma$ ($95.45$\% c.l.). For example, in the case of EGRET and COMPTEL experiments onboard \emph{CGRO}, where the statistics was almost always very low, it was not so unusual to read $2\sigma$ results. However, in X rays, with more statistics than in $\gamma$ rays, $3\sigma$ is generally an acceptable value for known sources, while in case of new detections, it is worth pushing to greater values up to $5-6\sigma$. Now, with the advent of \emph{Fermi Gamma-ray Space Telescope} with its exceptional performance, it is possible to search for $3$ or $5\sigma$ results also in the $\gamma-$ray energy band.

In high-energy astrophysics, as well as in other physical sciences, models are fitted to data and to estimate the goodness of a fit there are basically two tests. When a large number of counts is at hands (which, in turn, means to have a highly sensitive detector and/or a sufficiently high source flux), it is possible to apply the $\bf{\chi^2}$ \textbf{test}. This test can be used on binned data, with at least $20-30$ counts per bin. For example, let's consider a X-ray source emitting at a rate of $0.1$~c/s over the $1-10$~keV energy band. We would like to observe it for sufficient time to have a basic spectrum to be fitted with simple models (e.g. power-law). We want at least 10 bins over the whole energy range and, to apply the $\chi^2$ test, we need at least $20$ counts per bin. This means that we need a total of $20\times 10 = 200$ counts, which in turn can be obtained with $counts/rate=200/0.1=2000$~s of exposure. Obviously, there is surely some Murphy's law, according to which the source rate will decrease to $10^{-3}$~c/s just during your observation...

The goodness of the fit is estimated by the value of the reduced $\chi^2$ (generally indicated with $\tilde{\chi}^2$ or $\chi_{\nu}^2$). The $\tilde{\chi}^2$ is the ratio of $\chi^2$ divided by the number of degrees of freedom ($\nu$ or simply $d.o.f.$), which in turn is calculated by subtracting the number of constraints from the model to the number of bins. The fit is good when $\tilde{\chi}^2\leq 1$, while has to be rejected when $\tilde{\chi}^2>>1$. 

When there are no sufficient counts to generate a meaningful number of bins, because of either the Murphy's law above or the source is intrinsically weak for the performance of your instrument, it is no more possible to apply the $\chi^2$ test. In this case, it is necessary to use the \textbf{maximum likelihood}\footnote{In \texttt{xspec}, the fit of low counts spectra can be evaluated with the Cash statistic (command: \texttt{statistic cstat}), which is indeed the maximum likelihood (see Cash 1979).}, which is the joint probability that each single event can fit the adopted model (therefore, there is no need to bin the counts). That is, for each event, it is calculated the probability to fit the model and then, the product of all the probabilities is the likelihood function. By finding the maximum of this function, it is possible to estimate the parameters of the adopted model. The main disadvantage of the likelihood is that it is not possible to estimate the quality of the fit, since the maximum of the likelihood function just indicates the probability to obtain a specific observed result, but does not give any information on the expected probability. Some people use to perform Montecarlo simulations to bypass this issue, but the best that can be obtained is just a confirmation of the original measurement. Indeed, the original estimation of the parameters, used to set up the simulation, is obtained with likelihood and not with $\chi^2$. If you set up a Montecarlo simulation based on parameters estimated with likelihood and if these parameters are wrong, then the simulation does not change them by using a spell and, therefore, they remain wrong.   

It is possible to perform a likelihood ratio between the null  hypothesis $L_0$ (there is no source) and an alternative one, with likelihood $L_1$, corresponding to the fact that there is a source emitting with the specified model. The test statistic $TS = -2(\ln L_0 - \ln L_1)$ is related to the significance of the detection by the fact that $\sqrt{TS}$ is the number of $\sigma$ (i.e. $TS = 25$ corresponds to a $5\sigma$ detection). 

In addition to the statistical errors, \textbf{systematic errors} are often present in the processing of data and the calibration of astrophysical instruments. How to face and correct these errors is often left to the good will and initiative of the researcher, particularly at the beginning of a space mission, when the Instrument Teams themselves have not yet studied in detail the in-orbit performance of their instruments. The main problems with systematic errors are that they are generally of unknown origin during the earliest stages of the mission, often correlated, even in a complex way, so that it is not possible to extract a simple formula to use in the calculations. At high energies, the background can dominate in some energy bands and interact with the structures of the satellite or the instrument itself, generating secondary radiation that can be detected by the instrument (false triggers). Many other examples can be done. 

Therefore, in order to take into account these elements, it is often used a generic systematic error when fitting the data. In the famous tool for the analysis of X-ray spectra, \texttt{xspec}, there is the command \texttt{systematic}, which allows to add a systematic error in percent. However, this command is applied to all the loaded data, without any distinction, thus making the things even worse. This is particularly problematic when a fit on data from different instruments is done.

\begin{figure}
\centering
\includegraphics[scale=0.5]{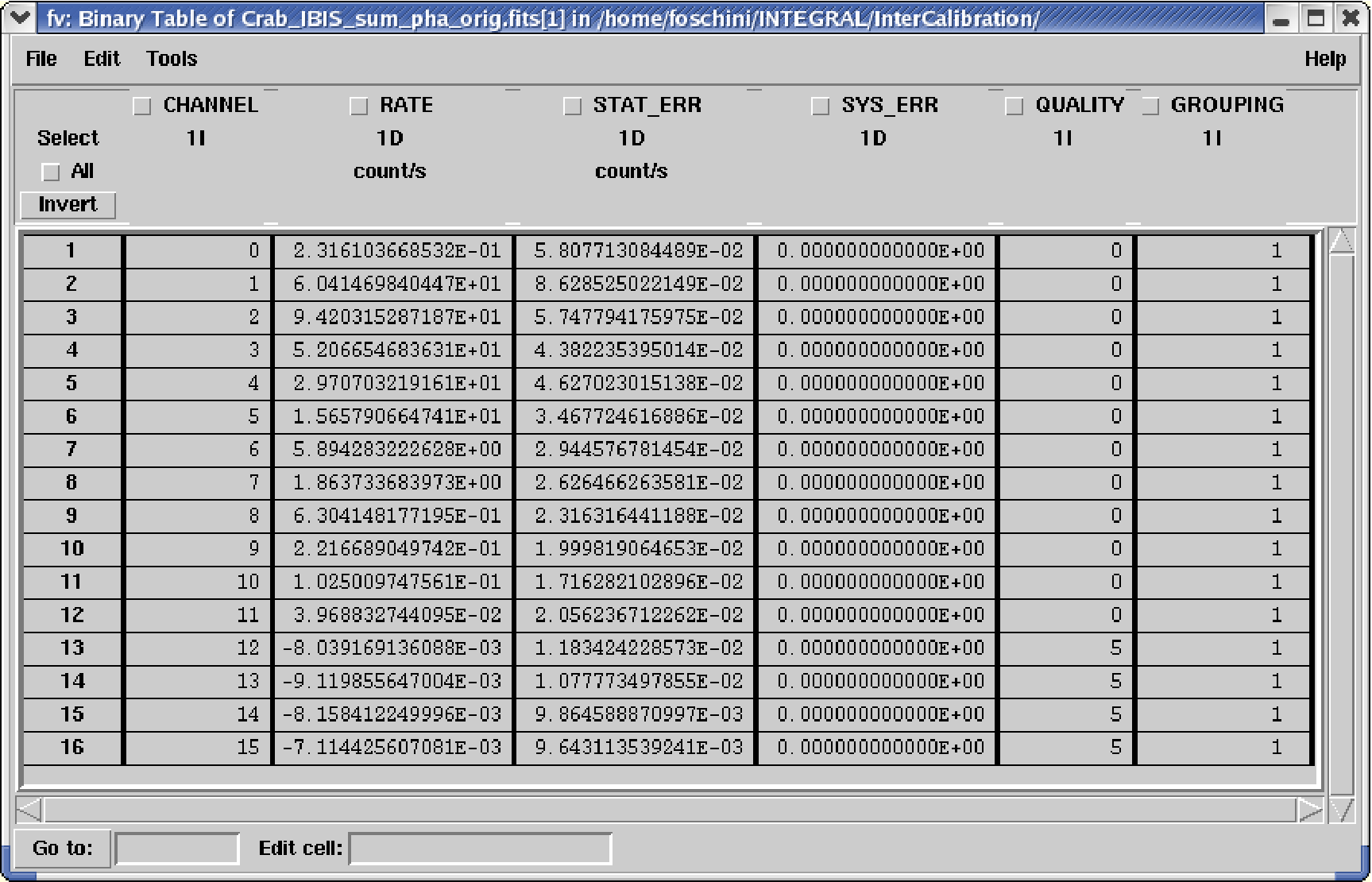}
\caption{Typical data structure of a count spectrum. In this case, the source is the Crab as observed by IBIS/ISGRI.}
\label{fig:syserr}
\end{figure}

The best option is to search and, when possible, to isolate and quantify the different effects that can cause the systematic errors. Then, it is possible to add the error in the energy band of interest directly in the data structure of the count spectrum. In Fig.~\ref{fig:syserr}, there is an example of the data structure of the spectrum extracted from IBIS/ISGRI data: the systematic error should be added in the proper row of the column \texttt{SYS\_ERR}.

Simulations can be a very useful tool to understand the origin of systematics and to estimate the amount of their values. The \texttt{xspec} command \texttt{fakeit} allows to perform some easy simulations, based on the of the instrumental response (RMF and ARF files). The Montecarlo models of the whole detector and spacecraft can give more detailed information, but are generally used by the Instrument Teams and not at hands of the individual scientist. A word of caution is worth saying, since simulations are sometimes used \emph{a bit too much}... Simulations are of paramount importance in the design and early calibration of the instrument, but please remind that are \emph{simulations}. Nothing can substitute the direct knowledge the instrument and to work on it. Therefore, if you measure a 10$\sigma$ effect, you do not need to run a Montecarlo simulation to know if it is real.... (and if a 10$\sigma$ effect is due to a malfunction of your instrument, it is better for you to switch it off and throw it in the outer space...).

The study of the \textbf{background} is so complex that can be a research topic by itself. The success (or not) of a space mission or, simpler, the analysis of data of your observation can strongly depend on this knowledge. Therefore, even the generic user must have more or less an idea on what data could be held, what should be thrown away, and how this operation can have impact on the final result.

The background can be generally divided into two parts:

\begin{itemize}
\item internal (\textbf{instrumental}) background, due to several elements; for example, the re-emission by fluorescence of the cosmic radiation impinging on the structures of the satellite or the instrument (like the tungsten of the IBIS mask that has a fluorescence line at $60-80$~keV, in the range of the low energy detector ISGRI);

\item the cosmic background, in its widest meaning, thus including the \textbf{Galactic} and the \textbf{Extragalactic} components; the former is thought to be due to the interaction of cosmic rays with the interstellar matter and is emitting mostly in the $\gamma-$ray band (MeV-GeV); the latter is though to be mainly due to non-resolved emission from Active Galactic Nuclei (AGN) and has a peak in the hard X rays ($\approx 30$~keV). 
\end{itemize}

Both of them are today very important research fields.

The \textbf{selection of the orbit} of the spacecraft is very important with respect to the impact of the background on the observations. In the Low-Earth Orbit (LEO, $400-600$~km), the satellites are shielded by the Earth magnetosphere (van Allen's belts), except when the spacecraft is passing through the South Atlantic Anomaly (SAA), a zone with anomalously low magnetic field, which allows the cosmic rays to reach the LEO altitudes. In the elliptical orbits (apogee $\approx 10^5$~km), the satellite exits from the van Allen's belts and therefore is less shielded. On the other hand, the elliptical orbits with great major axes are required in order to have long uninterrupted observing periods.

Last, but not least, there is one \textbf{instrumental bias} that cannot be avoided. Indeed, each instrument can always offer a partial view of any cosmic source, because it can measure the photons in a limited energy range, with a limited energy resolution and with some error in the coordinates reconstruction. There are certain sources that can best viewed in specific energy ranges and not in others. Remind that once there were the radio-selected and X-ray selected blazars. Now, we know that they are both members of the same family of cosmic sources, but this was not so straightforward at the beginnig. The only way to by-pass this selection bias is to perform multiwavelength observations. Perhaps, when in the future we will have also reliable and highly sensitive detectors for astroparticles (neutrinos, cosmic rays, ...), gravitational waves, and who knows what other type of ``messenger'', only at that time we could hope to have a really unbiased view of the cosmic sources.

\vskip 12pt
\textbf{To know more:}

\begin{itemize}
\item One chapter of the textbook by W.R. Leo, ``Techniques for Nuclear and Particle Physics Experiments'' (Springer-Verlag, 1994) is dedicated to the error analysis. Other textbooks, widely used, are: 
\subitem J.R. Taylor: ``An introduction to error analysis'' (University Science Books, 1997);
\subitem P.R. Bevington \& D.K. Robinson: ``Data reduction and error analysis for the physical sciences'' (McGraw-Hill, 2003).

\item There is a couple of papers dealing with the likelihood: 
\subitem W. Cash, ``Parameter estimation in astronomy through application of the likelihood ratio'', \emph{The Astrophysical Journal}, \textbf{228}, (1979), 939-947; 
\subitem J.R. Mattox et al., ``The likelihood analysis of EGRET data'', \emph{The Astrophysical Journal}, \textbf{461}, (1996), 396-407.

\item Recently was born the \emph{International Astronomical Consortium for High-Energy Calibration} (IACHEC) that should deal with all the issues of the cross-calibration of satellites for high-energy astrophysics are related problems. More details and documents can be found at:\
\texttt{http://www.iachec.org/}

\item A very good review on the $\gamma-$ray background is: 
\subitem A.J. Dean et al.: ``The modelling of background noise in astronomical gamma-ray telescopes'', \emph{Space Science Reviews}, \textbf{105}, (2003), 285-376.
\end{itemize}

\newpage

\section{The calibration sources at high energies}
In order to calibrate an instrument, it is necessary to have a stable and reliable source. The first source to be adopted for this purpose in high-energy astrophysics was the Crab. Its ``standard'' spectrum in the X-ray energy band is that measured by Toor \& Seward in 1974, which is a power-law model with $\Gamma=2.10\pm 0.03$ and normalization at $1$~keV equal to $9.7$~ph~cm$^{-2}$~s$^{-1}$~keV$^{-1}$. Recently, Kirsch et al. (2005) obtained consistent results by averaging the data from several instruments. 

During the earliest phases of the in-orbit life of an instrument (the performance verification phase), when the response is not yet fully ready, it is common use to measure the flux of cosmic sources in terms of Crab units and its fraction. For example, if an instrument operating in the $20-40$~keV band collects $100$~c/s from the Crab and it observes another cosmic source with a flux of $0.1$~c/s in the same energy band, then it is said that the source has a flux equal to $1$~milliCrab, which in turn is converted into physical units by means of the standard spectrum and corresponds to $7.7\times 10^{-12}$~erg~cm$^{-2}$~s$^{-1}$.

At $\gamma$ rays, another pulsar -- Vela -- is adopted as calibration source, because of its very high flux (of the order of $10^{-5}$~ph~cm$^{-2}$~s$^{-1}$ for $E>100$~MeV). As the energy increases, it is more and more difficult to find a useful calibration source.

During the latest years, the debate on the reliability of the Crab is becoming hotter and hotter. Specifically, thanks to the enormous technological improvements in the X-ray band (think, for example, to \emph{Chandra}), what was a point-like source is now a spatially resolved ensemble of different contributions. Moreover, the sensitivity of the instruments onboard \emph{Chandra} and \emph{XMM-Newton} has reached so high levels that the flux of the Crab is no more bearable and pile-up problems are rising. These occur when the flux is so high that two or more photons arrives within the same frame time of the detector (or readout time), so that they are recorded as one single photon with greater energy. This causes a distortion of the PSF and of the spectrum of the source (there is an excess of counts at high energies with a deficit of counts at low energies). This effect is more significant as the incoming flux increases. With the advancement of the performances of the instruments, it will be more and more difficult to use again the Crab as calibration source and already today there are ongoing campaigns dedicated to find other sources for this aim.

\vskip 12pt
\textbf{To know more:}

\begin{itemize}
\item The reference article for the Crab is:
\subitem A. Toor \& F.D. Seward, ``The Crab Nebula as a calibration source for X-ray astronomy'', \emph{The Astronomical Journal}, \textbf{79}, (1974), 995-999. 

\item An update with the data from modern satellites has been done by: 
\subitem M. Kirsch et al.: ``Crab: the standard X-ray candle with all (modern) X-ray satellites'', \emph{Proceedings of SPIE: UV, X-Ray, and Gamma-Ray Space Instrumentation for Astronomy XIV}, Edited by Siegmund, Oswald H. W., Volume 5898, pp. 22-33 (2005) [\texttt{astro-ph/0508235}].

\item The calibration at $\gamma$ rays is done by using the Vela pulsar. See, e.g.,
\subitem Abdo et al.: ``Fermi Large Area Telescope Observations of the Vela Pulsar'', \emph{The Astrophysical Journal}, \textbf{696}, (2009), 1084-1093.

\item Obviously, a lot of information and contacts can be found in the ``International Astronomical Consortium for High-Energy Calibration'' (IACHEC) site:\
\texttt{http://www.iachec.org/}
\end{itemize}

\newpage
\section{Final recommendations}
Someone once told that it is useless to give advices: wise people do not \emph{need} them, while silly persons do not \emph{use} them. Perhaps, I am a dreamer, but I think that it is still useful to give advices and, moreover, good advices are never sufficient. 

These few notes can just be the starting point for the data analysis in high-energy astrophysics and obviously there is a huge amount of things still to study and to say. However, I think that these notes can be a well-grounded basis and the references can give a reliable way to more details. For what remains, I do not see any other possibility than giving some general advices.

First of all, I would like to cite Henri Poincar\'e, who said: 

\begin{center}
\emph{Science is built up with facts, as a house is with stones.\\But a collection of facts is no more a science than a heap of stones is a house}.
\end{center}

The second thing worth reminding is the first (and perhaps the only) law of the software: 

\begin{center}
\emph{garbage in $\rightarrow$ garbage out}. 
\end{center}

That is, if you put garbage into your computer, then you can obtain only more garbage.

Last, but not least, in the uncertain field of the unknown, please remind that \emph{nobody} can tell you what is right and what is wrong.

\end{document}